\documentclass[12pt,preprint]{aastex}
\usepackage{color}

\newcommand{\vB}{{\bf B}}
\newcommand{\vJ}{{\bf J}}
\newcommand{\vv}{{\bf v}}

\newcommand{\um}{{\,\mu\rm m}}
\newcommand{\cm}{{\rm\,cm}}

\newcommand{\au}{{\rm\,AU}}
\newcommand{\AU}{{\au}}

\newcommand{\gm}{{\rm\,g}}
\newcommand{\gram}{{\gm}}
\newcommand{\kms}{{\rm\,km\,s^{-1}}}
\newcommand{\cms}{{\rm\,cm\,s^{-1}}}

\newcommand{\msun}{{\rm\,M_\odot}}

\newcommand{\second}{{\rm\,s}}
\newcommand{\muG}{{\,\mu\rm G}}

\newcommand{\gauss}{{\rm\,G}}
\newcommand{\Qunit}{{\cm^2\second^{-1}\gauss^{-1}}}

\begin{document}
\title{Disk Formation in Magnetized Clouds Enabled by the Hall Effect}

\author{Ruben Krasnopolsky\altaffilmark{1,2}, Zhi-Yun Li\altaffilmark{3,2}, Hsien Shang\altaffilmark{1,2}}
\altaffiltext{1}{Academia Sinica, Institute of Astronomy and Astrophysics, Taipei, Taiwan}
\altaffiltext{2}{Academia Sinica, Theoretical Institute for Advanced Research in Astrophysics, Taipei, Taiwan}
\altaffiltext{3}{University of Virginia, Astronomy Department, Charlottesville, USA}

\shortauthors{{\sc Krasnopolsky, Li, and Shang}}
\shorttitle{{\sc Disk Formation Enabled by the Hall Effect}}

\begin{abstract}
Stars form in dense cores of molecular clouds that are observed to
be significantly magnetized. A dynamically important magnetic field
presents a significant obstacle to the formation of protostellar
disks. Recent studies have shown that magnetic braking is strong
enough to suppress the formation of rotationally supported disks
in the ideal MHD limit. Whether non-ideal MHD effects can enable
disk formation remains unsettled. We carry out a first study on
how disk formation in magnetic clouds is modified by the Hall
effect, the least explored of the three non-ideal MHD effects
in star formation (the other two being ambipolar diffusion and
Ohmic dissipation). For illustrative purposes, we consider a
simplified problem of a non-self-gravitating, magnetized envelope
collapsing onto a central protostar of fixed mass. We find that
the Hall effect can spin up the inner part of the collapsing flow
to Keplerian speed, producing a rotationally supported disk. The
disk is generated through a Hall-induced magnetic torque. Disk
formation occurs even when the envelope is initially non-rotating,
provided that the Hall coefficient is large enough. When the magnetic
field orientation is flipped, the direction of disk rotation is
reversed as well. The implication is that the Hall effect can
in principle produce both regularly rotating and counter-rotating
disks around protostars.
We conclude
that the Hall effect is an important factor to consider in studying
the angular momentum evolution of magnetized star formation in
general and disk formation in particular.

\end{abstract}
\keywords{accretion, accretion disks --- magnetic fields --- ISM:
  clouds --- stars: formation}

\section{Introduction}
\label{intro}

Disks are an integral part of star formation; they are the birthplace
of planets. How they form is a long-standing, unresolved problem.
A major difficulty is that their formation is greatly affected by
magnetic braking, which has been hard to quantify until recently.

There is now increasing theoretical evidence that magnetic braking
may suppress the formation of rotationally supported disks (RSDs
hereafter) in
dense cores magnetized to a realistic level, with dimensionless
mass-to-flux ratios $\lambda$ of a few to several \citep{tc2008}.
\citet{als2003} first demonstrated through
2D (axisymmetric) simulations that RSDs
are suppressed by a moderately strong magnetic field in the
ideal MHD limit. \citet{glsa2006} showed analytically
that the disk suppression is due to the formation of a split
magnetic monopole, which is an unavoidable consequence of flux
freezing. The efficient disk braking was confirmed numerically
by \citet{ml2008} and \citet{hf2008} (see,
however, \citealt{mim2010} and \citealt{hc2009} for a
different view, and \citealt{lks2011} for a more detailed discussion).
The ideal MHD approximation must break down in order for RSD to form.

Two non-ideal MHD effects have already been explored in the context
of disk formation: ambipolar diffusion and Ohmic dissipation.
\citet{kk2002} found that ambipolar diffusion tends
to make disk braking more efficient, because it enables the
magnetic flux that would have gone into the central protostellar
object (and form a split monopole) in the ideal MHD limit to pile
up at small radii outside of the object instead, making the region
strongly magnetized (\citealt{lm1996}, \citealt{ck1998}, \citealt{cck1998}).
\citet{ml2009} showed
that the enhanced braking is strong enough to suppress the formation
of RSDs for realistic core conditions. The effect of Ohmic dissipation
was examined by several groups, starting with \citet{sglc2006}.
They suggested that enhanced resistivity (well above the classical
value) is needed for Ohmic dissipation to weaken the magnetic
braking enough to form a relatively large RSD of tens of AUs or
more. The suggestion was confirmed by \citeauthor{kls2010} (\citeyear{kls2010};  KLS10 hereafter),
although \citet{mim2010} were apparently
able to form RSDs using the classical resistivity computed by \citet{nnu2002}
(see also \citealt{db2010}). In any case, a third
non-ideal MHD effect, the Hall effect, has not been explored in
the context of disk formation (see, however, the independent
work of \citeauthor{b2011} in her unpublished PhD thesis); it is the
focus of this paper.

Our goal is to determine whether the Hall effect by itself can enable
an RSD to form in the presence of a relatively strong magnetic field
and, if yes, to estimate the magnitude of Hall coefficient needed
for RSD formation. We find that the Hall effect can actively spin up
the inner part of the protostellar collapsing flow, potentially to
Keplerian speed, unlike the other two non-ideal MHD effects. The
combined effect of all these three non-ideal MHD terms on disk formation
will be explored in another investigation (\citealt{lks2011}; LKS11 hereafter).

\section{Problem Setup}
\label{setup}

We adopt the same problem setup as in KLS10, where we consider the
collapse onto a star of $0.5\msun$ of a rotating envelope
that is uniformly magnetized initially. In the absence of any
non-ideal MHD effect, the magnetic field would prevent an RSD from
forming through magnetic braking. The Hall effect changes the
evolution of the magnetic field (and thus the braking efficiency)
through the induction equation:
\begin{equation}
\frac{\partial \vB}{\partial t}= \nabla (\vv\times \vB)
  - \nabla \times \left\{ Q [(\nabla \times \vB)\times {\bf
	B}]\right\}
 \label{induction}
\end{equation}
where $Q$ is a coefficient of the Hall effect which, for simplicity,
we will assume to be spatially constant. The MHD equations are solved
using ZeusTW, a 3D non-ideal MHD code based on Zeus3D \citep{cnf1994}.
We treat the Hall term in the induction equation using
an explicit method based on \citet{ss2002} and \citet{h2003},
which includes subcycling to speed up the computation.

As in KLS10, we adopt a spherical polar
coordinate system ($r$, $\theta$, $\phi$), and fill the computation
domain with an isothermal gas of sound speed $a=0.2\kms$
between $r_i=1.5\times 10^{14}\cm$ and $r_o=1.5\times 10^{17}\cm$.
A uniform density $\rho_0=1.4\times 10^{-19}\gram\cm^{-3}$ is
assumed, so that the total envelope mass is $1\msun$.
For the initial rotation, we adopt the following prescription:
\begin{equation}
v_\phi = v_{\phi,0} \tanh(\varpi/\varpi_c)
\label{rotation}
\end{equation}
where $\varpi$ is the cylindrical radius, and $v_{\phi,0}=2\times
10^4\cm\second^{-1}$ and $\varpi_c=3\times 10^{15}\cm$ are chosen.
With this setup, an RSD of $\sim 400\AU$ in
radius is produced at a representative time of $t=10^{12}\second$ in
the absence of any magnetic field (see Fig.\ 1 of KLS10). It is
completely destroyed by a moderately strong initial magnetic
field of $B_0=35.4\muG$ in the ideal MHD limit (see Fig.\ 2 of
KLS10). Whether the RSD can be restored by the Hall effect is the
question that we seek to address.

\section{Result}
\label{result}

\begin{figure}
\epsscale{1.25}
\plottwo{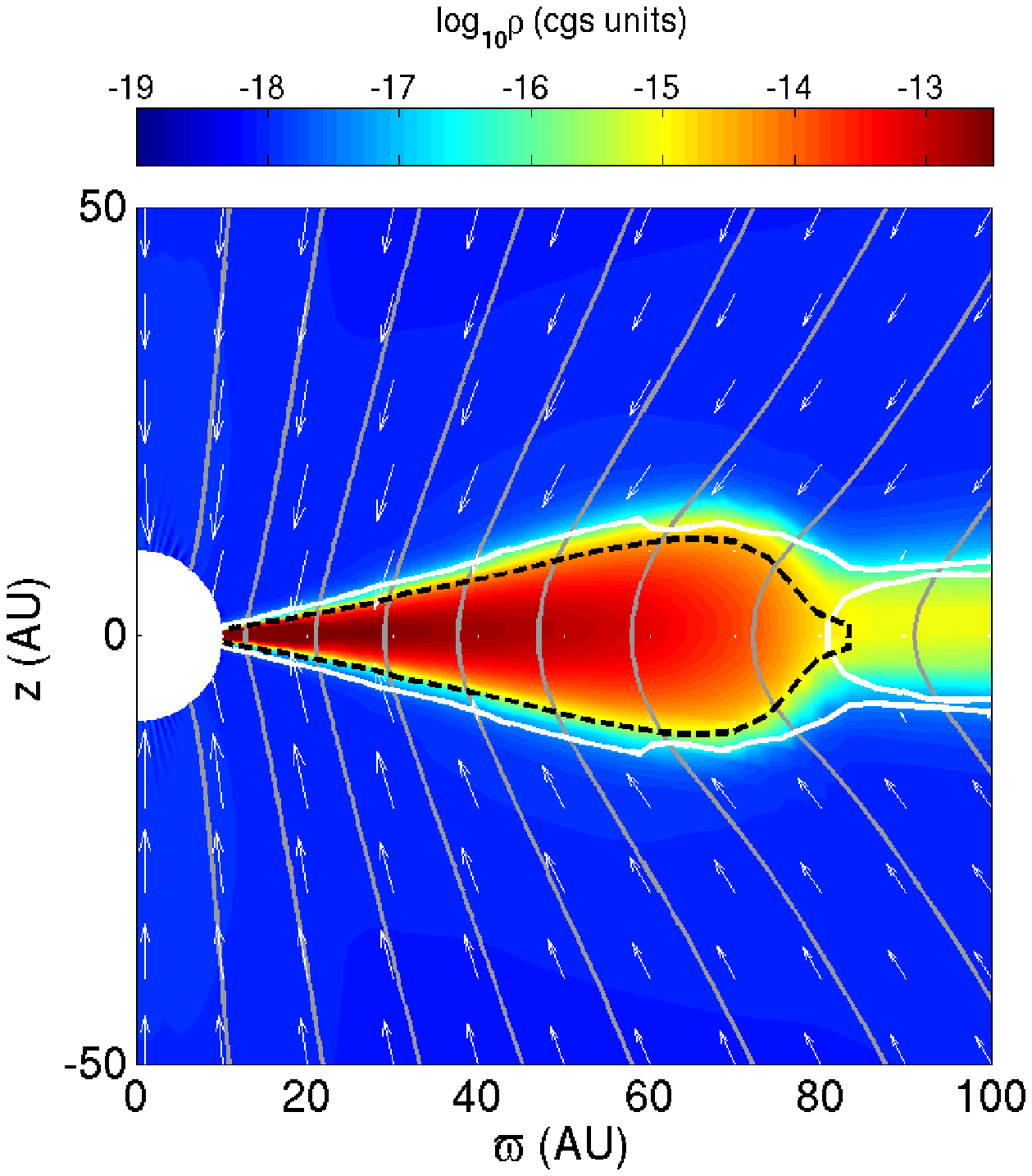}{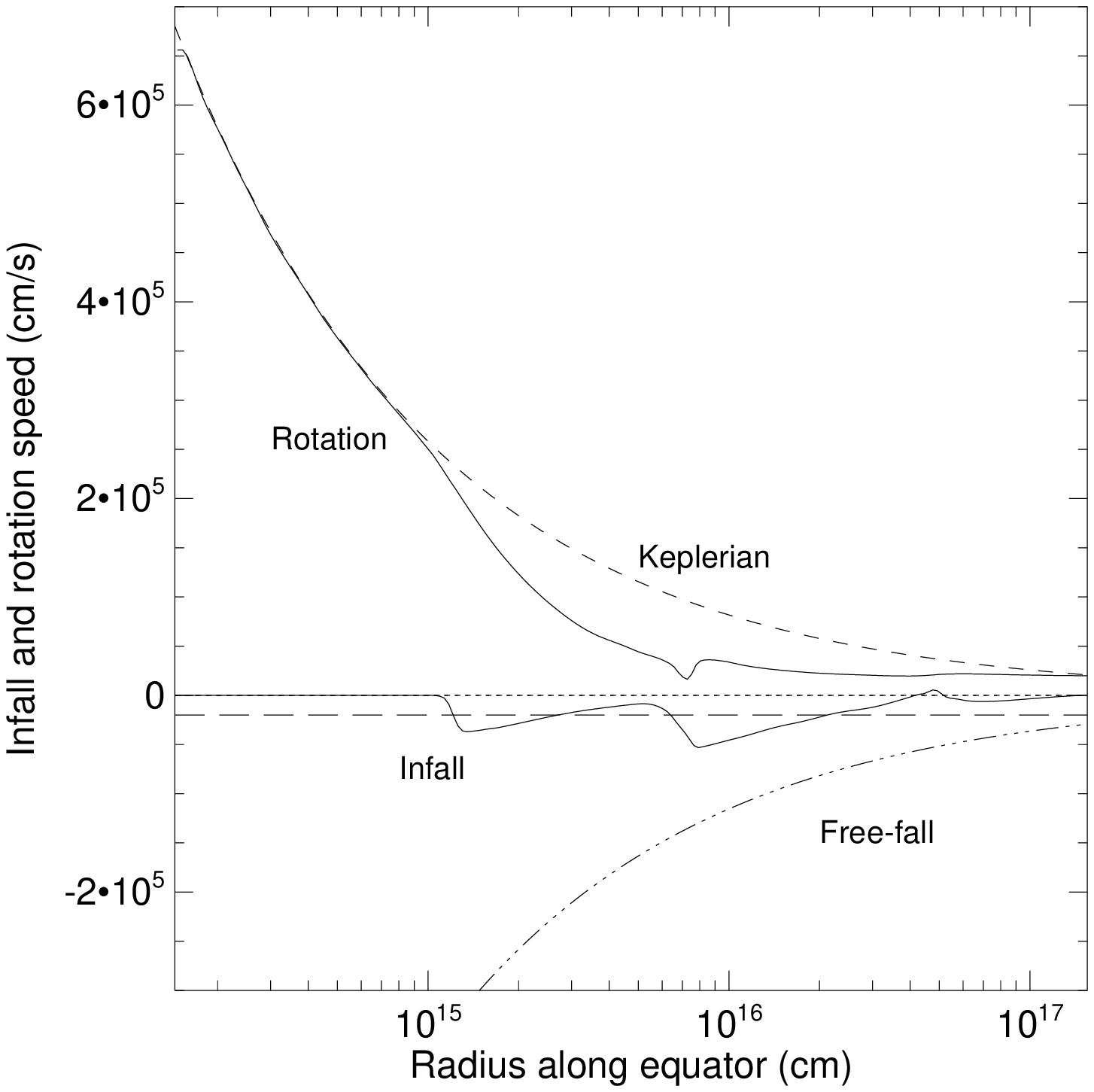}
\caption{Left: Snapshot of the rotationally supported disk formed
through Hall effect in the meridian plane at a representative
time $t=2\times 10^{12}\second$ for Model PARA. Logarithm of density
(color map); radial infall sonic transition
$v_r=-a=-2\times 10^4\cms$ (white line);
poloidal velocity field (arrows); magnetic field lines (gray solid lines);
level of Keplerian support $|v_\phi/v_K|$
(dark gray dashes at 90\%),
where $v_K=(G M_\ast/\varpi)^{1/2}$. Right: infall and
rotation speeds on the equator. Also plotted for
comparison are the Keplerian
speed and free fall speed, the sound speed (horizontal dashed
line), and zero speed line.
}
\label{hall_19}
\end{figure}

\begin{figure}
\epsscale{1.25}
\plottwo{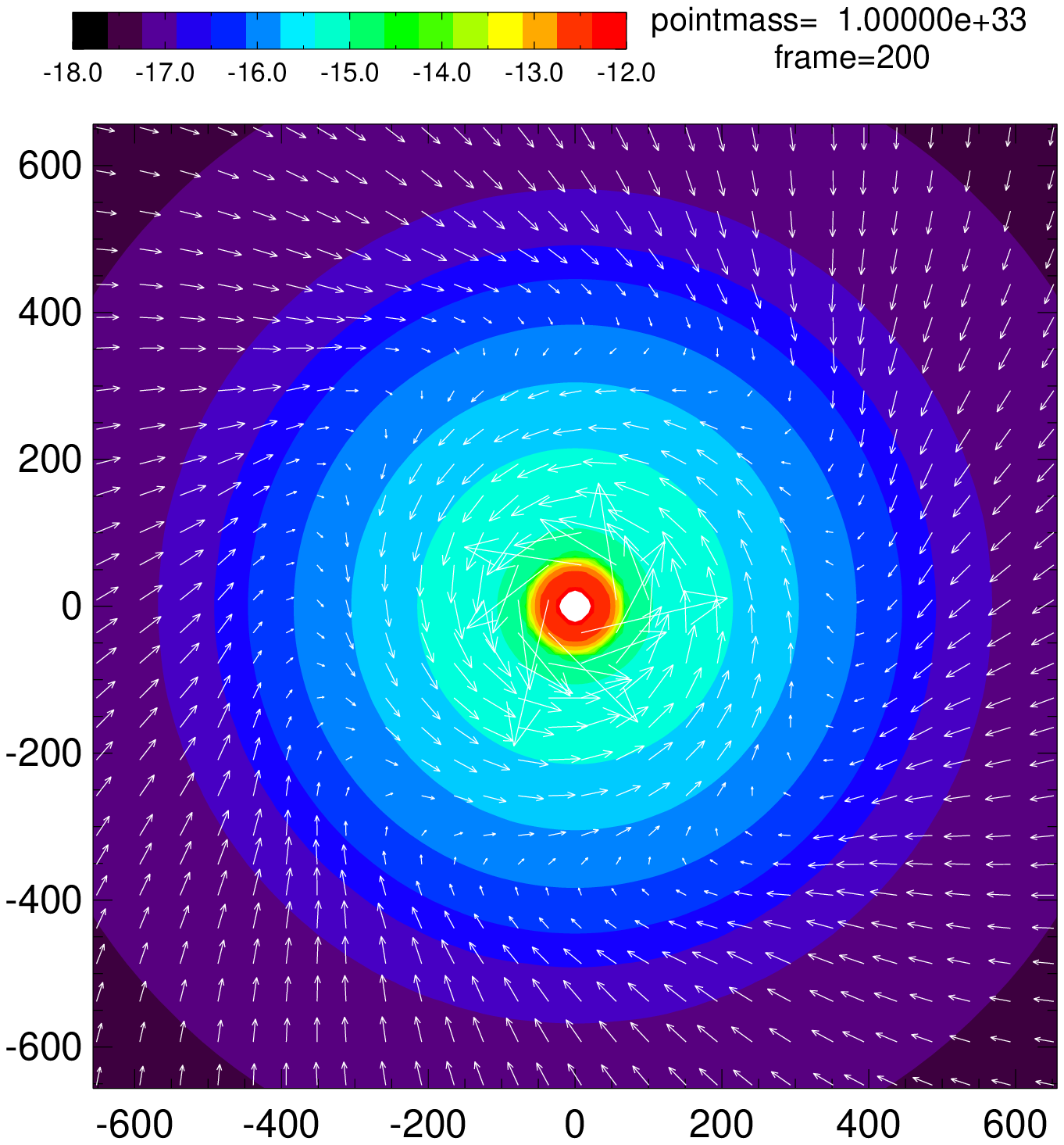}{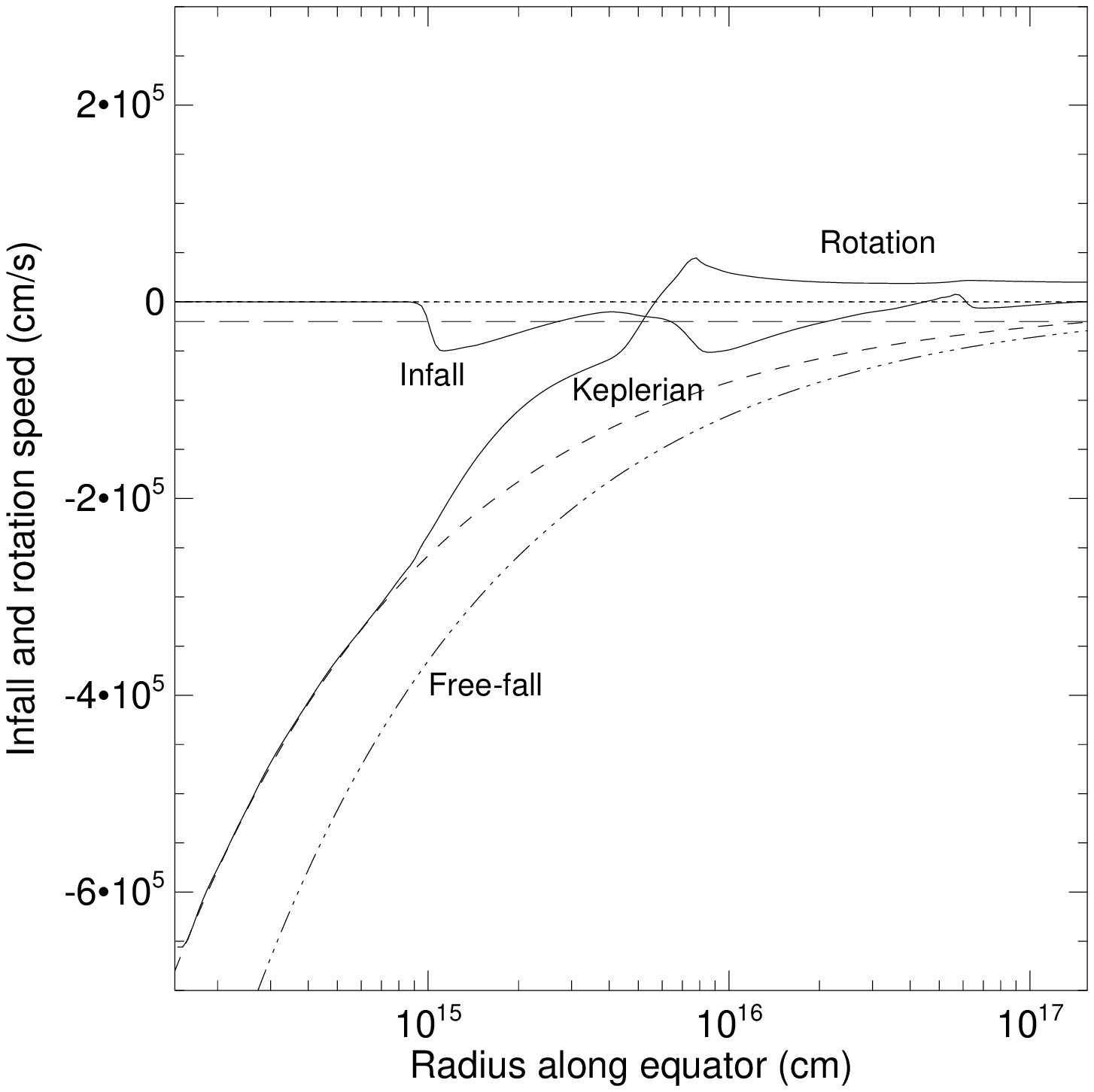}
\caption{Left: Snapshot of the rotationally supported disk formed
through Hall effect in the equatorial plane at a representative
time $t=2\times 10^{12}\second$ in the equatorial plane, with
anti-aligned initial magnetic field and rotation axis (Model ANTI).
Note that the Hall effect forces the inner part rotate in a direction
opposite to that of the outer part. Right: infall and rotation speeds
on the equator, with the Keplerian and free fall speeds also
plotted for comparison.
}
\label{hall_22}
\end{figure}

We consider first a model where the initial magnetic field is parallel
to the rotation axis (Model PARA in Table \ref{table:first}) and the
coefficient $Q^\prime =
3\times 10^{22}$ in the Lorentz-Heaviside units convenient to use with the Zeus family
of codes, corresponding to $Q=Q^\prime/\sqrt{4\pi}=8.46\times 10^{21}$
in Gaussian CGS units ($\Qunit$). An RSD is able to
form in this case, as
illustrated Fig.\ \ref{hall_19}, which plots a color map of density
distribution, velocity field and magnetic field lines in the meridian
plane (left panel), and
the infall and rotation speeds on the equator (right panel), at a
representative time $t=2\times 10^{12}\second$. The RSD shows up clearly
on the color map as the flattened high density region of about
$60\AU$ in radius. It has an infall speed that is close to zero
and a rotation speed that is Keplerian. The Hall effect has clearly
enabled an RSD to form, although it is not by weakening the magnetic
braking that would have suppressed the RSD formation in the ideal
MHD limit, unlike the case of Ohmic dissipation (KLS10). The RSD
is formed because the inner part of the collapsing flow is actively
spun up by the Hall effect. A strong support for Hall spin-up
as the cause for the RSD formation comes from a second simulation
where we flip the sign of the initial magnetic field direction
(Model ANTI), so that the field is anti-parallel, rather than parallel,
to the rotation axis. In this case, an RSD disk is again formed, but
it rotates in a direction opposite to the initial rotation, as shown
in Fig.\ \ref{hall_22}. The reversal of the disk rotation direction
means that the Hall effect does not merely weaken the magnetic
braking so that enough of the original angular momentum is retained
for RSD formation.

\begin{deluxetable}{lllll}
\tablecolumns{6}
\tablecaption{Model Parameters \label{table:first}}
\tablehead{
\colhead{Model} & Q (cgs units) & \colhead{ $B_0$ ($\mu{\rm G}$) }  &
\colhead{Rotation}  & \colhead{RSD?}
}
\startdata
PARA  & $8.46 \times 10^{21}$ & 35.4  & yes  &  yes  \\
ANTI  &$8.46 \times 10^{21}$  & - 35.4  & yes  &  yes   \\
NoROT   & $8.46 \times 10^{21}$  & 35.4  & no &  yes  \\
LoB   & $8.46 \times 10^{21}$  & 10.6  & no  &  yes  \\
LoQ   & $8.46 \times 10^{20}$  & 35.4  & no  &  yes   \\
LoQ1   & $2.82 \times 10^{20}$  & 35.4  & no  &  yes   \\
LoQ2   & $8.46 \times 10^{19}$  & 35.4  & no  &  no   \\
\enddata
\end{deluxetable}

The reason that the disk rotation direction is reversed when the
magnetic field direction is flipped is the following (see \citealt{wk1993}
for a similar discussion in the context of disk-wind
launching). From the induction
equation (\ref{induction}), it is easy to see that the Hall effect
forces the field lines to move with a velocity
\begin{equation}
\vv_{\rm H} = - Q (\nabla \times \vB) = - {\frac{4\pi Q}{c}} \vJ
\label{HallDrift}
\end{equation}
where $\vJ$ is the current density. The pinching of the poloidal
magnetic field near the equatorial region produces a strong toroidal
current, $j_\phi$, which changes sign as the poloidal field direction
is flipped. The toroidal current of different signs causes the field
lines to wind up toroidally in opposite directions, giving rise to
magnetic torques of opposite sense. In the case where the initial
field line is parallel to the rotation axis, $j_\phi$ is positive,
giving rise to a toroidal field-line speed $v_{{\rm H},\phi}$ that is
negative (assuming that the Hall coefficient is positive, which is
not necessarily true, see, e.g., \citealt{wn1999} and LKS11;
$v_{{\rm H},\phi}$ changes sign if $Q$ is negative). Since the toroidal
current is the strongest on the equator, the field line is twisted
backward most on the equator and less so above and below the
equatorial plane. The differential twist forces the field line to
bend in the positive azimuthal direction, creating a torque that
spins up the material near the equatorial plane in the same
direction as the initial rotation. This explains why the RSD in
the field-rotation aligned case rotates in the same sense as the
initial rotation. When the initial field direction is flipped, the
toroidal current $j_\phi$ (and thus $v_{{\rm H},\phi}$) changes sign,
forcing the field lines to bend in the negative azimuthal
direction. The resulting negative torque not only overcame the
initial positive rotation but also produced a counter-rotating RSD
in the anti-aligned case.

\begin{figure}
\epsscale{1.25}
\plottwo{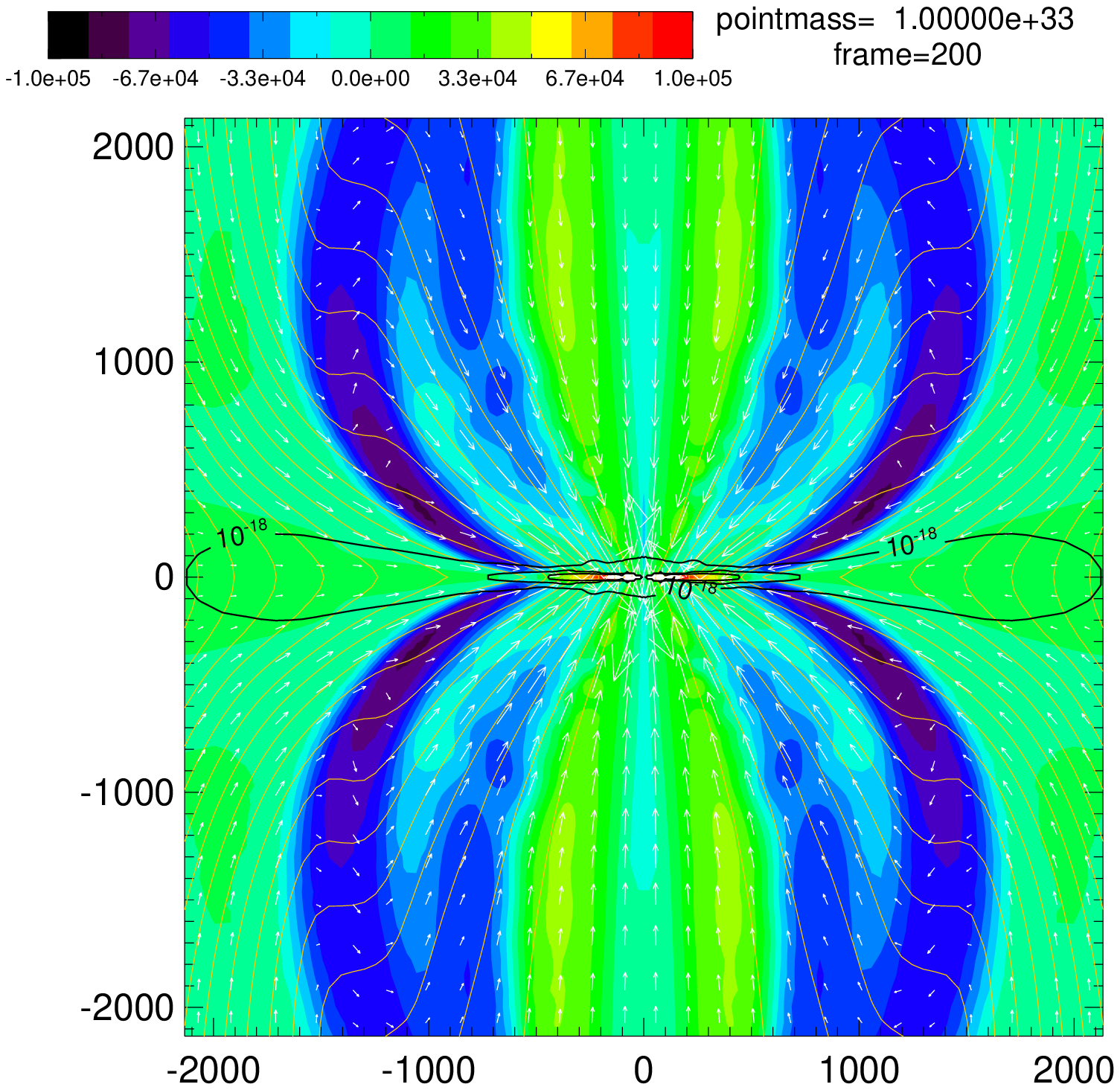}{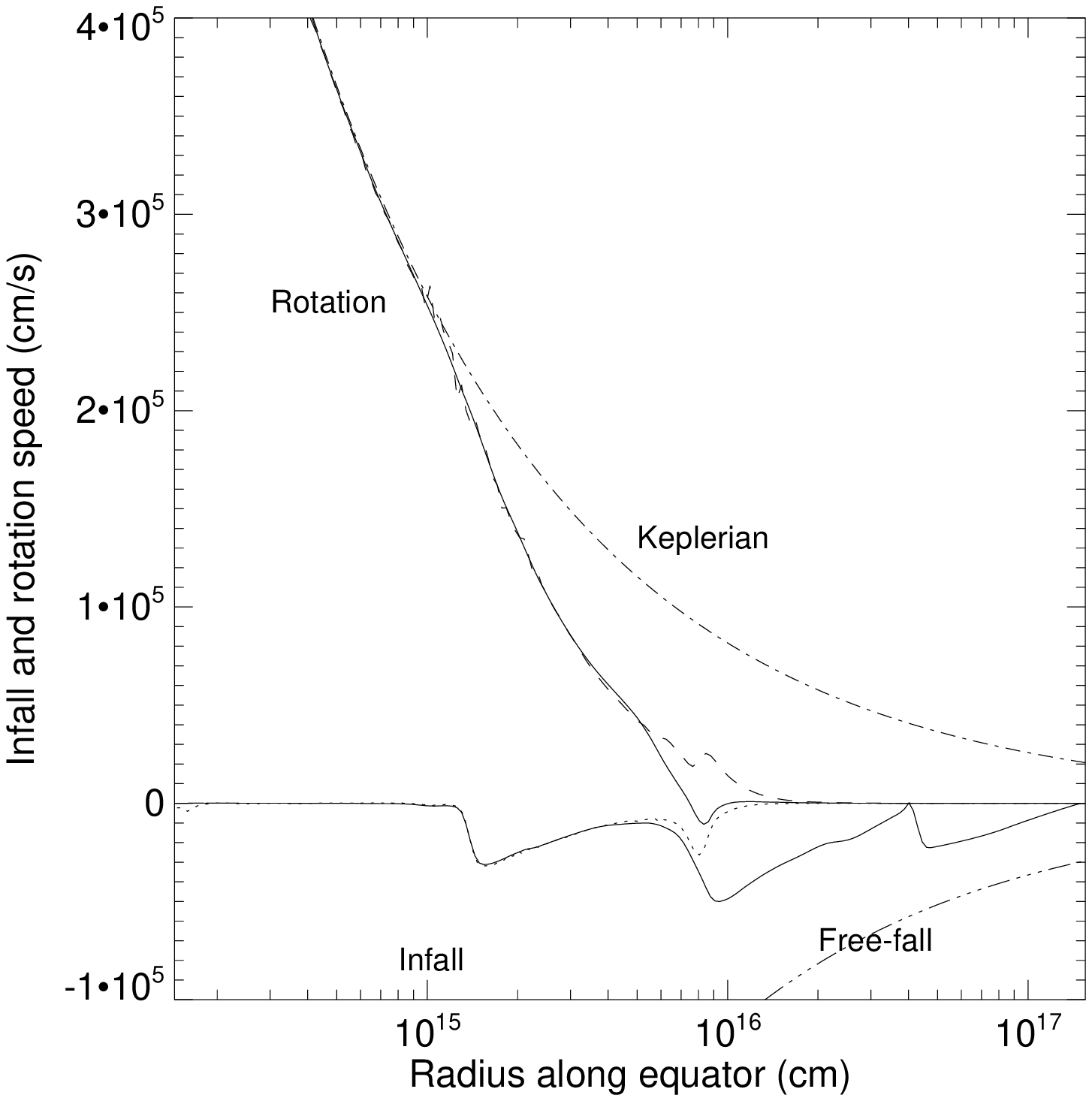}
\caption{Left panel: map of the Hall-induced rotation speed in
an initially non-rotating envelope at a representative
time $t=2\times 10^{12}\second$ (Model NoROT; the unit for length
is AU and for speed in the colorbar is cm/s). Also plotted are
magnetic field lines (yellow), isodensity contours (black contours, in steps
of 10), and velocity vectors (white). Right panel: infall and rotation
speeds on the equator. Also plotted are the negative of the
radial and toroidal component of the Hall-induced field line drift
velocity, $-v_{{\rm H},r}$ (dotted line) and $-v_{{\rm H},\phi}$ (dashed).
}
\label{hall_24}
\end{figure}

The Hall spin-up is even more unambiguous in Model NoROT where the
envelope is initially non-rotating; any
subsequent rotation must be
due to the Hall effect. The right panel of Fig.\ \ref{hall_24} shows
that an RSD is again formed in this case, with the rotation
speed approaching the Keplerian speed and the infall speed dropping
close to zero inside a radius $\sim 10^{15}\cm$. Since the envelope
is initially non-rotating, the spin of the RSD must be balanced by
material rotating in the opposite direction. The left panel of
Fig.\ \ref{hall_24} demonstrates that this is indeed the case. It
shows that, outside the equatorial plane, materials of positive
and negative angular momentum occupy alternating shells. Inside
a radius of $\sim 5\times 10^2\AU$, the angular momentum is
positive, dominated by the (small) RSD and highly flattened
pseudodisk outside RSD (see the isodensity contours). At larger
distances, the angular momentum becomes negative, dominated by
the outermost counter-rotating shell shown in the Fig.\ \ref{hall_24}.
The total positive angular momentum ($1.73\times
10^{52}\gram\cm^2\second^{-1}$) does not exactly balance out the total
negative angular momentum ($-2.12 \times 10^{52}\gram\cm^2\second^{-1}$),
however, indicating some angular momentum is transported out of
the simulation box, presumably through torsional Alfv\'en waves.
Note that the Hall effect modifies the envelope rotation well
beyond the $10^2\AU$-scale RSD.
We have verified that when the initial magnetic field direction
is flipped, the direction of the Hall-induced rotation is reversed.

Besides torquing up the envelope in the toroidal direction, the
Hall effect also enables the magnetic field lines to diffuse out
radially. In the absence of any radial diffusion, the field lines
would be pulled by the accreting material into a split magnetic
monopole, which is not present in any of the simulations discussed
so far. Indeed, at small radii where the Hall spin-up is most
efficient, the inward advection of poloidal magnetic field is
almost exactly balanced by the outward magnetic diffusion
enabled by the Hall effect. This balance is illustrated in the
right panel of Fig.\ \ref{hall_24}, which shows that, inside a
radius $\sim 10^{16}\cm$, the radial component of the Hall-induced
field line drift velocity on the equator has almost the same
magnitude as the infall speed but with an opposite sign. The
radial magnetic diffusion allows material to fall inward
without dragging field lines along with it. The situation is
similar in the azimuthal direction, where the Hall-induced
field line drift speed and the fluid rotation speed are
nearly identical in magnitude but opposite in sign. The
azimuthal Hall-induced drift prevents the field lines from
being wound up continuously by rotation. The required toroidal current
is provided by the bending of the field lines inside the disk
(see left panel of Fig.\ 1; contrast with the unbent field lines
in Fig.\ 7 of KLS10).

An interesting feature of the envelope collapse is that the infall
speed of the equatorial material remains well below the free-fall
speed. The equatorial collapse is
slowed down three times (see right panel of Fig.\ \ref{hall_24}),
for different reasons. The first slowdown near $\sim 4\times 10^{16}\cm$
corresponds to the edge of the magnetic bubble inflated by
field twisting (outside the region shown in the right panel of
Fig.\ \ref{hall_24}), where a magnetic
barrier forces the collapsing material over a large solid angle into
a narrow equatorial channel (see Fig.\ 2 of \citealt{ml2008} and
associated discussion). The second slowdown occurs near $\sim
8\times 10^{15}\cm$ (or about $500\AU$), inside which the accreted
magnetic flux is left behind due to Hall diffusion. It is similar
in origin to the ambipolar diffusion-induced accretion shock
first discussed in \citeauthor{lm1996} (\citeyear{lm1996}; a similar behavior is also
present in the Ohmic dissipation-only case, LKS11). The increase
in the poloidal magnetic field strength enabled by Hall diffusion
in the radial direction makes it easier to spin up the equatorial
material through the Hall-induced magnetic torque. The third
slowdown near $\sim 10^{15}\cm$ is obviously due to centrifugal
effect, since an RSD is formed interior to it.

\begin{figure}
\epsscale{1.25}
\plottwo{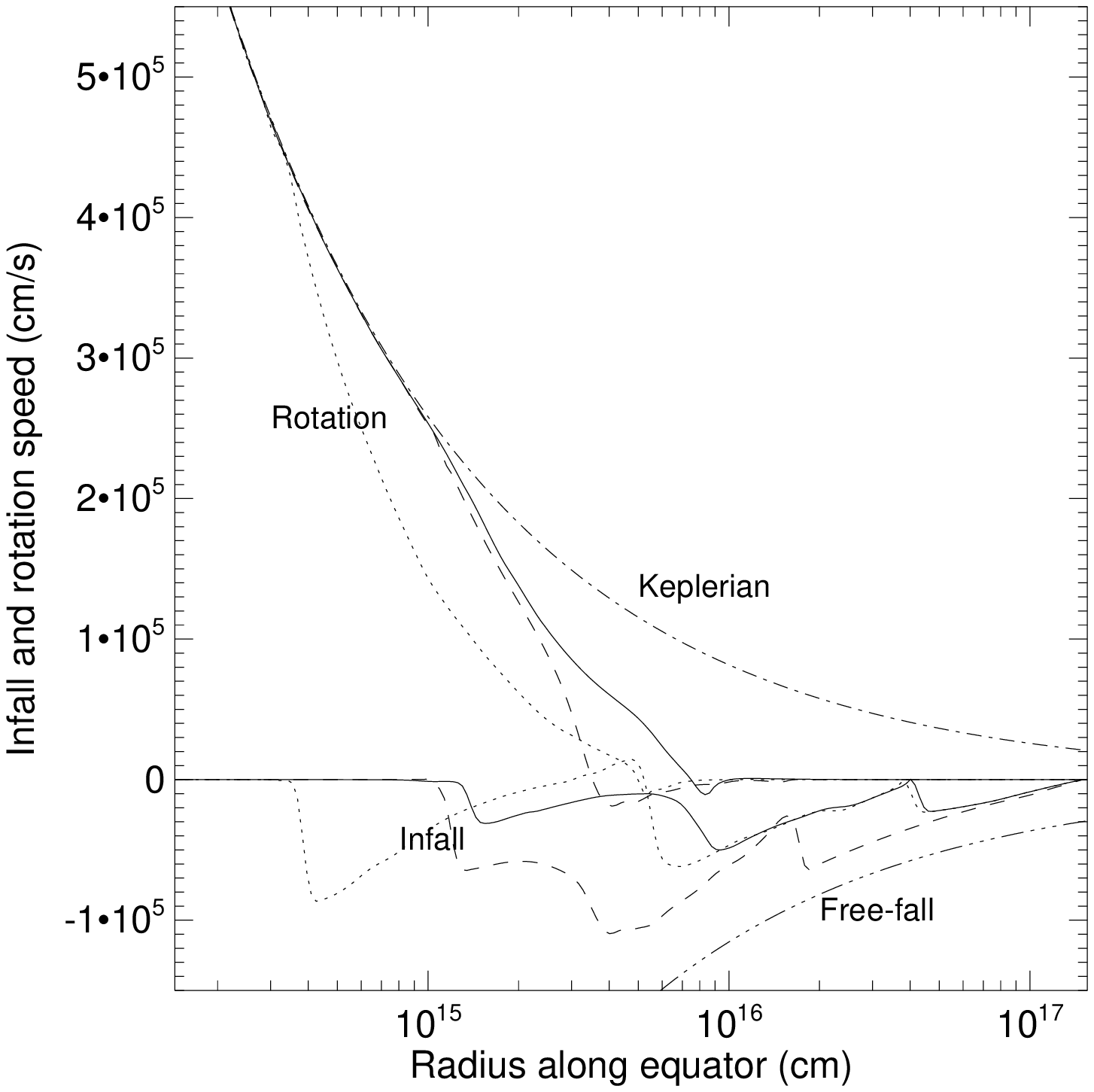}{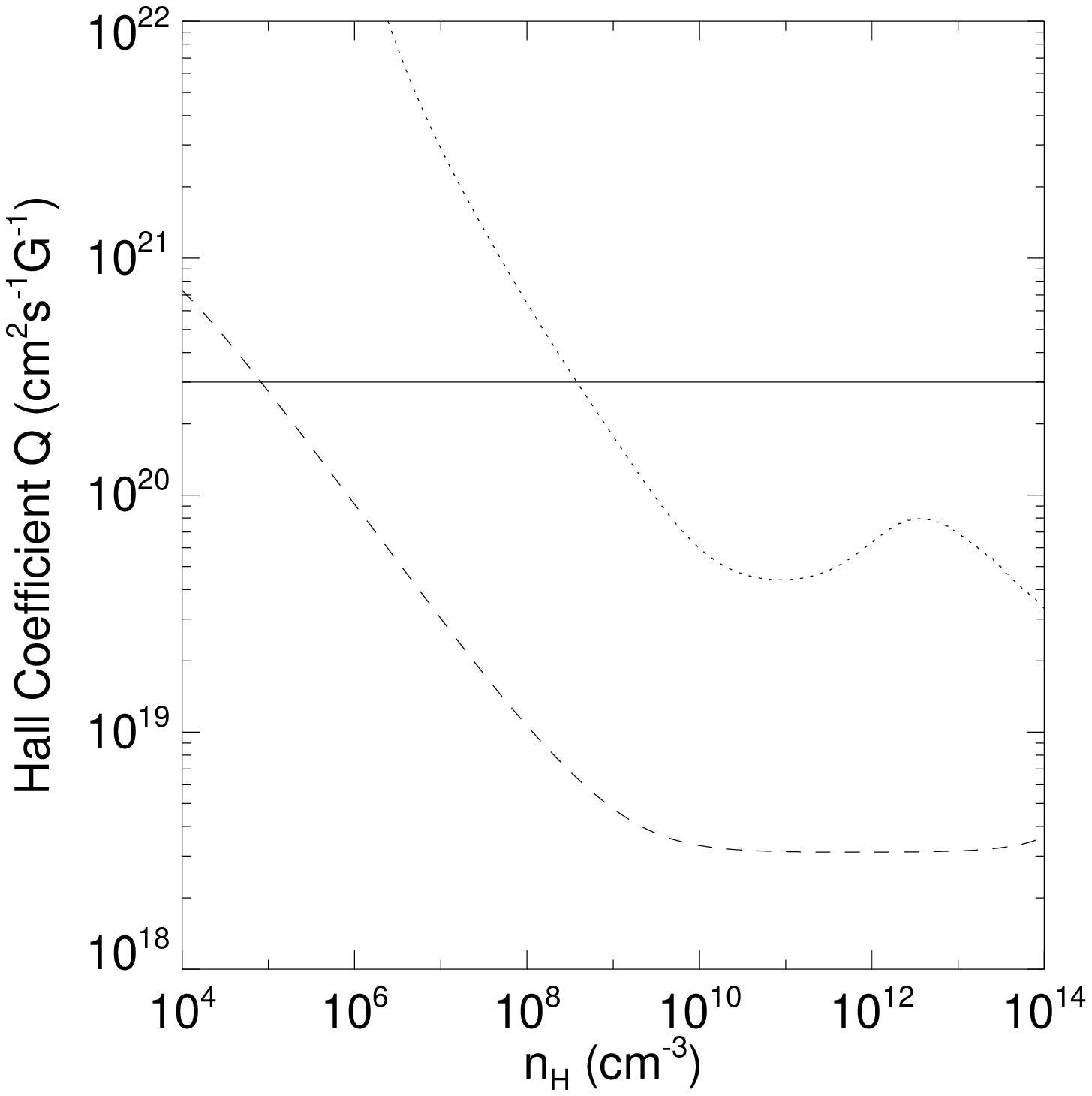}
\caption{Left: Comparison of infall and rotation speeds on the equator
for the non-rotating Model NoROT (solid lines) and its two
variants: Models LoB (dashed) and Model LoQ (dotted).
Right: Comparison of the estimated critical value for the Hall
coefficient $Q_c$ for RSD formation (solid line) with the microscopic
values of $|Q|$ computed for the MRN grain size distribution (dotted)
and $1\um$ sized grains (dashed), assuming a cosmic ray ionization
rate of $10^{-17}\second^{-1}$.
}
\label{hall_compare}
\end{figure}

To check how robust the above results from Model NoROT are, we did a
number of variants of the model. First, we reduced the initial
field strength to $30\%$ of its original value, from $35.4\muG$ to
$10.6\muG$ (Model LoB). The result is shown in the left panel of
Fig.\ \ref{hall_compare} (see
dashed lines). An RSD is again formed at the representative time
$t=2\times 10^{12}\second$, despite the reduced initial field strength.
It is somewhat smaller than that in the original case, which is
to be expected since the Hall spin-up depends on current density,
which increases with field strength. Another difference is that
the infall speed outside the RSD is higher in the weaker field
case, because the magnetic forces that oppose gravity are weaker.
Second, we reduced the coefficient $Q$ by a factor of 10, from
$8.46\times 10^{21}$ to $8.46\times 10^{20}$ in Gaussian cgs units (Model
LoQ). An
RSD is still formed at $t=2\times 10^{12}\second$. With a radius of
$\sim 3\times
10^{14}\cm$ or $20\AU$, it is much smaller than the RSD in the
original Model NoROT. If we reduce the value of $Q$
by another factor of 3 (to $2.82\times
10^{20}\Qunit$,
Model LoQ1),
an RSD still forms, but it extends barely outside the inner
boundary of our computation domain, which has a radius of
$10\AU$. When the value of Q is
reduced further to $8.46\times 10^{19}\Qunit$
(Model LoQ2), the RSD disappears. We therefore take
$3\times 10^{20}\Qunit$ as a rough
estimate for the value of $Q$ needed for the Hall-enabled
formation of a relatively large RSD of tens of AUs or more
in size.

We have verified that the RSD formed in Model NoROT is little
affected when the classical Ohmic resistivity used in KLS10
(based on \citealt{nnu2002}) is included. An enhanced
resistivity can, however, reduce the size of the RSD
significantly. Experimentation shows that it suppresses the
Hall-enabled RSD formation in Model NoROT altogether when
the resistivity is on the order of $10^{20}\cm^2\second^{-1}$
or larger. The negative effect of the resistivity on the
Hall spin-up is not surprising, since it reduces the electric
current density in general, and the toroidal current
density in particular; the latter is the driver of the
spin-up through the Hall-induced magnetic torque.

How does the critical value of $Q_c \sim 3\times
10^{20}\Qunit$ estimated for
RSD formation compare with the microscopic values of expected in
realistic dense cores? In a separate study (LKS11), we have computed
the values of $Q$ as a function of hydrogen number density
$n_{\rm H}=\rho/(2.33\times 10^{-24}\gram$), based on a
simple prescription for the magnetic field strength as a function
of density \citep{nnu2002} and a simplified chemical network
for charge densities \citep{nnu1991}, including two extreme
grain size distributions: (1) the standard MRN power-law distribution
(\citealt{mrn1977}; appropriate for diffuse interstellar clouds),
and (2) grains of single, large size of $1\um$. These two cases
should bracket the situation in dense molecular cloud cores, where
some grain growth is expected. As can be seen from the right panel
of Fig.\ \ref{hall_compare}, although the microscopic values of $Q$
can be larger than the critical value $Q_c$ at relatively low
densities, they are smaller than $Q_c$ by about an order of
magnitude at densities of order $n_{\rm H}\sim 10^{10}\cm^{-3}$
(or higher) that are crucial for RSD formation (see the left
panel of Fig.\ \ref{hall_19}).

\section{Conclusion and Discussion}
\label{discussion}

We have studied the collapse of a non-self-gravitating rotating,
magnetized envelope onto a central stellar object of fixed mass
to illustrate the influence of the Hall effect on disk formation.
In this idealized problem, the formation of a rotationally
supported disk (RSD) is completely suppressed by magnetic
braking in the ideal MHD limit. Including a Hall term with
a coefficient $Q \gtrsim 3\times 10^{20}\Qunit$ in the
induction equation enables an RSD of radius $\gtrsim 10\AU$ to
form. The RSD is formed not because the Hall effect has reduced
the efficiency of the magnetic braking. Rather, it is produced
because the Hall effect actively spins up the inner part of the
equatorial material to Keplerian speed. The spin-up comes about
because the collapsing envelope drags the magnetic field into
a highly pinched configuration near the equator, producing a
large toroidal electric current density, which forces the field
lines to rotate differentially due to the Hall effect. The
resulting twist of field lines yields the magnetic torque that
spins up the equatorial material, even in the absence of any
initial rotation. The spin-up is most effective in the inner part
of the accretion flow where radial Hall diffusion enables the
magnetic flux that would have been dragged into the central
object by the accreted material in the ideal MHD limit to
stay behind; the resulting pileup of magnetic flux at small
radii is similar to the cases with either ambipolar diffusion
or Ohmic dissipation. When the field direction is
flipped, the Hall-induced magnetic torque changes direction
as well. An implication is that the direction of the
angular momentum of the material close to a protostar,
including RSD, may depend more on the magnetic field
orientation than on the initial rotation of the dense
core that the star is formed out of, at least in principle.

In practice, it is uncertain whether the Hall-induced magnetic
torque can produce a sizable RSD or not. First, the value of the
coefficient $Q$ required to produce an RSD of tens of AUs in size
in our idealized model appears to be larger than the microscopic
values expected in dense cores, by roughly an order of
magnitude. Whether the coefficient can be enhanced somehow,
perhaps through anomalous processes, is unclear. More
importantly, the Hall effect is only one of the non-ideal MHD
effects that are present in the lightly ionized, magnetized,
dense cores. It can be dominated by ambipolar diffusion at
low densities and by Ohmic dissipation at high densities.
It will be interesting to explore the interplay of all these three
non-ideal MHD effects and how they affect the
collapse of dense cores and the formation of protostellar disks.
Nevertheless, we have demonstrated that the Hall effect
is unique among the non-ideal effects in its ability to actively
spin up a magnetized collapsing flow, potentially to Keplerian
speed, providing a new mechanism for disk formation.

\acknowledgments
We acknowledge support by the Theoretical Institute for
Advanced Research in Astrophysics (TIARA), by the
National Science Council of Taiwan through grant
NSC97-2112-M-001-018-MY3, and by NASA through NNG06GJ33G
and NNX10AH30G.

\end{document}